\documentclass[twocolumn]{aastex631}
\usepackage{amsmath}
\usepackage{gensymb}
\usepackage{comment}
\usepackage{bm}
\usepackage{array}

\begin{document}

\title{Automated Scheduling of Doppler Exoplanet Observations at Keck Observatory}

\author[0000-0002-9305-5101]{Luke B. Handley}
\affiliation{Department of Physics \& Astronomy, University of California Los Angeles, Los Angeles, CA 90095, USA}

\author[0000-0003-0967-2893]{Erik A.\ Petigura}
\affiliation{Department of Physics \& Astronomy, University of California Los Angeles, Los Angeles, CA 90095, USA}

\author[[0000-0002-8952-5617]{Velibor V. Mi\v{s}i\'{c}}
\affiliation{Anderson School of Management, University of California Los Angeles, Los Angeles, CA 90095, USA}

\author[0000-0001-8342-7736]{Jack Lubin}
\affiliation{Department of Physics \& Astronomy, University of California Los Angeles, Los Angeles, CA 90095, USA}

\author[0000-0002-0531-1073]{Howard Isaacson}
\affiliation{501 Campbell Hall, University of California at Berkeley, Berkeley, CA 94720, USA}
\affiliation{Centre for Astrophysics, University of Southern Queensland, Toowoomba, QLD, Australia}
%confirmed



\begin{abstract}

Precise Doppler studies of extrasolar planets require fine-grained control of observational cadence, i.e. the timing of and spacing between observations. We present a novel framework for scheduling a set of Doppler campaigns with different cadence requirements at the W. M. Keck Observatory (WMKO). For a set of observing programs and allocated nights on an instrument, our software optimizes the timing and ordering of $\sim$1000 observations within a given observing semester. We achieve a near-optimal solution in real-time using a hierarchical Integer Linear Programming (ILP) framework. Our scheduling formulation optimizes over the roughly $10^{3000}$ possible orderings. A top level optimization finds the most regular sequence of allocated nights by which to observe each host star in the request catalog based on a frequency specified in the request. A second optimization scheme minimizes the slews and downtime of the instrument. We have assessed our algorithms performance with simulated data and with the real suite of Doppler observations of the California Planet Search in 2023.

\end{abstract}

\keywords{methods: observational}


\section{Introduction} \label{sec:intro}
The discovery and characterization of extrasolar planets with the Doppler technique requires careful attention to the observational `cadence' defined as the {\em timing of} and {\em spacing between} observations of a given target. The census of known Doppler extrasolar planets have orbital periods ranging from only a few hours to several decades \citep{Howard2012,Fulton2021}. Thus, Doppler surveys must be tuned to appropriately sample the orbital periods of interest. 

Moreover, time-variable surface features produce shifts in the stellar spectra that register as Doppler shifts, yet have nothing to do with the star’s motion around the star-planet barycenter \citep{Luhn2020}. The amplitude of activity RVs ranges from tens of cm/s for the very quietest stars to hundreds of m/s for young and active stars. The timescale of this stellar variability ranges from minutes (acoustic modes) \citep{Chaplin2019} to hours (granulation), to days (rotation), to years (magnetic cycles) \citep{Fulton2015}. 

Over the last few years, a number of spectrometers have been commissioned that are stable at the level of several tens of cm/s. Some noteworthy examples include ESPRESSO \citep{Pepe2021}, MAROON-X \citep{Seifahrt2018}, EXPRES \citep{Blackman2020}, NEID \citep{Schwab2016}, and KPF \citep{Gibson2020}. For such instruments, stellar activity is the dominant noise source for nearly all target stars. Several strategies have been developed for mitigating stellar activity. Most require dense sampling of RVs over timescales relevant to planetary and activity signals. Therefore, control of observational cadence has become even more critical as the community attempts to detect planetary signals below 1 m/s in the presence of stellar activity \citep{Anglada-Escude2016}. 

Achieving the desired observational cadence at any Doppler facility is a complex scheduling challenge. Most facilities support a number of active Doppler programs with different targets, observational priorities, and cadence requirements. In addition, most Doppler instruments share the telescope with other instruments with their own scheduling constraints (e.g. dark time for extra-galactic observations). Finally, weather losses are guaranteed but the set of impacted nights is not known in advance. 

Today, the task of scheduling Doppler observations is performed almost entirely by humans. This is a challenging and time-consuming task. Schedulers must determine the sequence of observations that favor high program completion and equity among the supported programs. There is an opportunity for automated scheduling algorithms to save human effort and better achieve completion/equity goals. At present, we are aware of three Doppler facilities that are autonomously scheduled: the Las Cumbres Observatory Network of Robotic Echelle Spectrographs (LCO/NRES, \citealt{Siverd18}), the Calar Alto high-Resolution search for M dwarfs with Exo- earths with Near-infrared and optical Echelle Spectrographs (CARMENES, \citealt{Quirrenbach14}), and the Automated Planet Finder Levy Spectrograph \citep{Vogt14}. 

Automated telescope scheduling has been deployed since at least the early 1990s. For a more detailed review of prior efforts, see \cite{Solar16}, \cite{Bellm19}, \cite{mushrooms}, and references therein. A key early development was the Hubble Space Telescope's SPIKE software \citep{Johnston94} which has been adapted to suit other facilities including the James Webb Space Telescope \citep{Giuliano08}. Many of the scheduling algorithms developed in the 2000s adopted `local search' or `greedy' algorithms. Here a schedule is built sequentially by selecting the next observation that contributes the most to a merit function. Such methods are not well suited to Doppler work because they lack the look-ahead needed to schedule observations over long intervals. 

A significant recent development was the introduction of Integer Linear Programming (ILP) methods to the telescope scheduling problem by \cite{Lampoudi15} who developed the dynamic scheduling algorithm for the LCO network. ILP and Mixed-Integer Linear Programing (MILP) formulations of scheduling problems are common in the operations research communities and powerful commercial software libraries exist to solve these problems to their global optima. The ability to confidently find global optima makes these methods promising for the long-range planning needs of Doppler surveys. ILP- or MILP-based schedulers have since been developed for the Atacama Large Millimeter/Submillimeter Array (ALMA) by \cite{Solar16} and the Zwicky Transient Facility (ZTF) by \cite{Bellm19b}.

This paper describes our efforts to automatically schedule Doppler observations at Keck Observatory. We provide some background on the prior human scheduling at Keck Observatory  (Section~\ref{sec:background}). We describe our scheduling algorithm (Section~\ref{sec:sem}). We present, apply and assess the performance of our scheduler during observations at Keck observatory during 2023 (Section~\ref{sec:implementation}). Finally, we conclude and offer some thoughts on future development (Section~\ref{sec:future}). 

\section{Manual Scheduling at Keck}
\label{sec:background}
Doppler searches for extrasolar planets have been conducted at the 10~m Keck-I telescope with the High Resolution Echelle Spectrometer (HIRES) instrument since 1994 \citep{Marcy1996}. The telescope is classically scheduled. PIs submit proposals to a number of time allocation committees (TACs) who award full or fractional nights over a six-month observing semester. Since 2010, the Doppler programs have received $\approx$40--50 nights per semester distributed over $\approx$60--70 full or fractional nights. Generally, awarded time has been anchored by a small number of large, multi-semester projects, such as NASA mission support of Kepler, K2 and TESS, and a larger number of smaller 1 to 3 night projects. The set of all accepted programs (Doppler and other) are sent to the Keck Observatory with a cover sheet that includes basic scheduling constraints and preferences. Those requests include a set of nearly consecutive nights and at least one night per month to monitor short and long period planetary orbits. A human scheduler at Keck observatory heuristically balances these requests to determine a full semester schedule for the telescope.

Most PIs of Doppler programs at Keck elect to execute their observations with the infrastructure of the California Planet Search (CPS, \citealt{Howard10}). CPS runs an {\em ad hoc} queue which schedules observations and maintains a pool of trained observers to execute the observations. CPS also reduces, curates, and distributes data. By collaborating with CPS, PIs may execute observations over a larger number of nights and share in weather losses. Indeed, many of smaller programs could not achieve scientifically useful cadence without access to a large number of nights through CPS.

Between 2010 and 2020, CPS observed $\approx$40,000 stellar spectra (not counting multi-shot exposures) or an average of 2000 observations per semester. Each semester's scheduling challenge is substantial as there are, in principle, $\mathcal{O}(N!) \sim 10^{6000}$ possible orderings. The CPS scheduler collects the observation requests at the start of the semester and before each night of observing. On the day preceding observations, the scheduler manually marks and removes previous observations. The schedule is balanced each night with fractions of each program as a function of overall allocations. This helps balance weather losses, especially when entire nights are lost and each program loses time proportionally. The scheduler must also balance the distribution of target right ascension (RA) so that the telescope is not over- or under-subscribed at any point in the night. Follow-up of exoplanet systems in the ecliptic further complicate the scheduling with the need to avoid pointing near the moon.

After the scheduler determines upcoming night's targets, they generate a script, an ordered sequence of observations to be executed at the telescope. The scheduler also minimizes overheads associated with long slews and cable wrap limits. This task is analogous to the traveling salesman problem with the additional complications of target visibility windows and time-dependent slew times. The scheduler must have an intuitive understanding of the relationship between the right ascension/declination coordinate system (that specify the targets) and the altitude/azimuth coordinate system (that specifies the telescope orientation). The relationship between these coordinate systems is a time-dependent and non-trivial transformation requiring tens of nights of observing experience to understand for planning purposes. The final schedule is slightly overfilled relative to what can be achieved even under exceptional observing conditions. At the telescope, CPS observers make real-time decisions to skip targets given current conditions and their knowledge of the scientific needs of the portfolio of programs.

The human effort required to schedule CPS observations is substantial. At the beginning of a semester, with the most targets to consider, the ratio of scheduling planning versus observation can be as high as one-to-one, and is reduced to one-to-ten by the end of the semester. PIs of contributing programs must also monitor the progress of their programs and request changes, if needed. By any standard, the amount of human time required to plan effectively while scheduling up to 100 stars in a night is high.

\section{Scheduling Algorithm} 
\label{sec:sem}

Determining the precise timestamps of 1000 exposures throughout every night in an observing semester is an impractical feat when both problem size and weather uncertainty are considered. Precise minute-to-minute scheduling is only necessary for an upcoming night or small sequence of nights. However, long range look-ahead at the semester scale remains critical to the scientific goals of the algorithm. To balance these needs, our scheduling algorithm is hierarchical, splitting the problem into two stages. 

First, it solves the scheduling problem at quarter night resolution, assembling groups of targets into each allocated quarter night while incorporating constraints on time, visibility, and cadence. The targets do not have a specified ordering within each quarter during this stage. In the second step, a small sequence of the nearest upcoming quarters is scheduled \textit{exactly} with the targets assigned by the first step. In this paper, we describe the long range semester scheduling stage in detail, including the initialization data, model construction, and objective function. The second step, which we refer to as the `Traveling Telescope Problem' or the `TTP', is described in detail in \cite{handleyTTP}, submitted to {\em AJ}.

\subsection{Inputs to the Scheduler} 
\label{sec:inputs}

We begin the semester scheduling with a list of $N_r$ many requests $\{1,\ldots,N_r\}$, where $r$ is the request index. Each request has the following attributes:

% NOTATION
% N_r - Number of Requests total
% n_r^inter - Number of nights visited
% n_r^intra - Exposures per night

\begin{itemize}
\item Star name {\em [string]}.
\item Celestial coordinates {\em [RA, Dec, epoch]}.
\item Maximum number of unique nights of observation over the semester {\em [integer]}, $n_r^\mathrm{inter}$
\item Maximum number of exposures per scheduled night {\em [integer]} $n_{r}^\mathrm{intra}$.
\item Minimum duration between observations on different nights in days, the {\em inter-night spacing} {\em [integer]}, $\tau_{r}^\mathrm{inter}$.
\item Minimum duration between observations on the same night in hours, the {\em intra-night spacing} {\em [float]} $\tau_{r}^\mathrm{intra}$.
\item Exposure time $\tau_{r}^\mathrm{exp}$ in seconds {\em [integer]}
\item Program code {\em [string]} that associates each request with an approved telescope program.
\end{itemize}
Table \ref{tab:inputs} shows a sample of KPF requests for the 2023B semester to illustrate the form of the input data.

\begin{table*}
\caption{Sample of KPF requests. From top to bottom: a non-cadenced/single shot observation, a low cadence observation, a medium cadenced observation, and a high cadenced observation with three observations requested per night.}
\label{tab:inputs}
\begin{tabular}{lccccccccc}
\hline
\text{Starname} & \text{Ra} & \text{Dec} & $n_r^\mathrm{inter}$ & $n_r^\mathrm{intra}$ & $\tau_{r}^\mathrm{inter}$ & $\tau_{r}^\mathrm{intra}$ & $\tau_{r}^\mathrm{exp}$ & \text{Program code} \\
& Hour & $\degree$ & & & Day & Hour & Sec & \\
\hline
KOI-4032 & 19.0 & 42.7 & 1 & 1 & \text{NaN} & \text{NaN} & 120 & IB \\
K00701 & 18.9 & 45.3 & 5 & 1 & 15 & \text{NaN} & 1200 & LW \\
K00117 & 19.8 & 48.2 & 20 & 1 & 1 & \text{NaN} & 1200 & LW\\
K00319 & 18.8 & 43.9 & 18 & 1 & 2 & \text{NaN} & 1400 & JZ \\
T006324 & 22.1 & 67.5 & 14 & 3 & 1 & 1.5 & 900 & FD \\
\hline
\end{tabular}
\end{table*}

The scheduler also requires a list of KPF quarter nights. If $N_{s}$ is the total number of these allocated quarters, we index the list of slots arranged by increasing time $\{1,\ldots,N_{s}\}$ with the letter $s$. Each individual awarded quarter night $s$ need not be adjacent in the Keck schedule (in fact they are rarely so). We denote the duration of each slot in seconds with $I_{s}$.

Finally, since the ILP algorithm described below partitions observations into slots while satisfying the cadence constraints, we need to specify which targets can be observed on which slots. This is encoded in the `Visibility Matrix' $V_{rs}$.
\begin{itemize}
    \item $V_{rs}$ = 1 if the target $r$ is visible during quarter night slot $s$, and 0 otherwise.
\end{itemize} 
During the 2023B semester, we calculated $V_{rs}$ directly from the target coordinates, following a simple set of rules described in Section~\ref{sec:visible}. However, $V_{rs}$ may be specified according to an arbitrary set of rules, such as maximum airmass or minimum moon distance. We constrain the scheduling of targets only to when they are visible in Section \ref{sec:semcon}.  

Finally, the schedule requires an array that maps each slot $s$ to a specific calendar date. This is done with $t_{s}$ which is set to be 1 on the first day of the observing semester. The time difference (in days) between two slots $s$ and $s'$ is simply $\delta = t_{s'}-t_{s}$. We denote the maximum separation of any two given nights as $\delta_\mathrm{max} = t_{N_{s}}-t_{1}$, the duration of the observing semester. We summarize the symbols from the main body of this text in Appendix~\ref{sec:variables}.

\subsection{Weather Sampling} \label{sec:weather}
Variable weather at Maunakea impacts the nightly schedules. Some nights, cloud cover or poor seeing leads to significant reductions of throughput, rendering our target list practically unachievable. On severe weather nights, we are forced to forfeit observing altogether. While we cannot predict the precise extent of weather-related disruptions in advance, failing to incorporate amortized weather losses into our scheduler would introduce a substantial error into our completion expectations.

To address these challenges, we reserve 30\% of all allocated quarter nights for adverse weather conditions. During these reserved periods, no observational targets are scheduled. This 30\% allocation applies to a random selection of time slots in $\{1, 2, \ldots, N_{s}\}$. In other words, for each slot $s'$ randomly designated for weather-related downtime, we ensure that all observing requests $r$ have $V_{rs'} = 0$.

We chose to sample entire quarters as opposed to sampling out 30\% of the time within every quarter night because we believe this to be closer to the reality of weather losses. In general, individual quarters are usually either lost entirely to weather, or not. This works as a first order correction to our forecasting, and we discuss how this might be improved in Section \ref{sec:future}.

\subsection{Decision Variables} \label{sec:semvar}

Our ILP formulation of the semester scheduling problem uses the following binary decision variables:

\begin{itemize}
\item $Y_{rs}$ = 1 if request $r$ is scheduled to the quarter night slot $s$, and 0 otherwise. This variable specifies the full semester schedule, with the caveat that requests within a slot are not (yet) ordered. It has size $N_r \times N_s$.

\item $B_{rss'}$ = 1 if requested target $r$ is scheduled to \textit{both} slots $s,s'$, and 0 otherwise. This variable tracks the assignment of pairs of observations. It has size $N_r \times N_s^2$.

\item $D_{r\delta}$ = 1 if requested target $r$ is observed twice with time separation $\delta$ in days, and 0 otherwise, tracking the cadence achieved for each request. It has size $N_r \times (\delta_\mathrm{max}+1)$.
\end{itemize}

\begin{figure}
    \centering
    \vspace{.3cm}
    \includegraphics[width=\columnwidth]{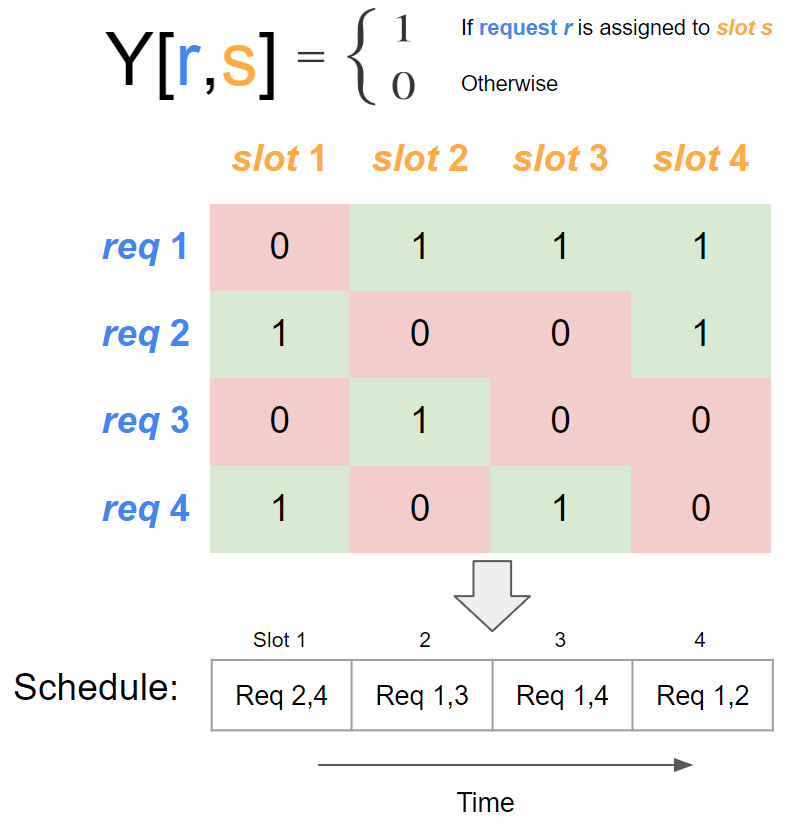}
    \caption{The $Y_{rs}$ matrix specifies the set of requests assigned to each slot. Note that a single request may be scheduled multiple times and a slot may contain multiple requests.}
    \label{fig:yvariable}
\end{figure}

\begin{figure*}
\includegraphics[width=\textwidth]{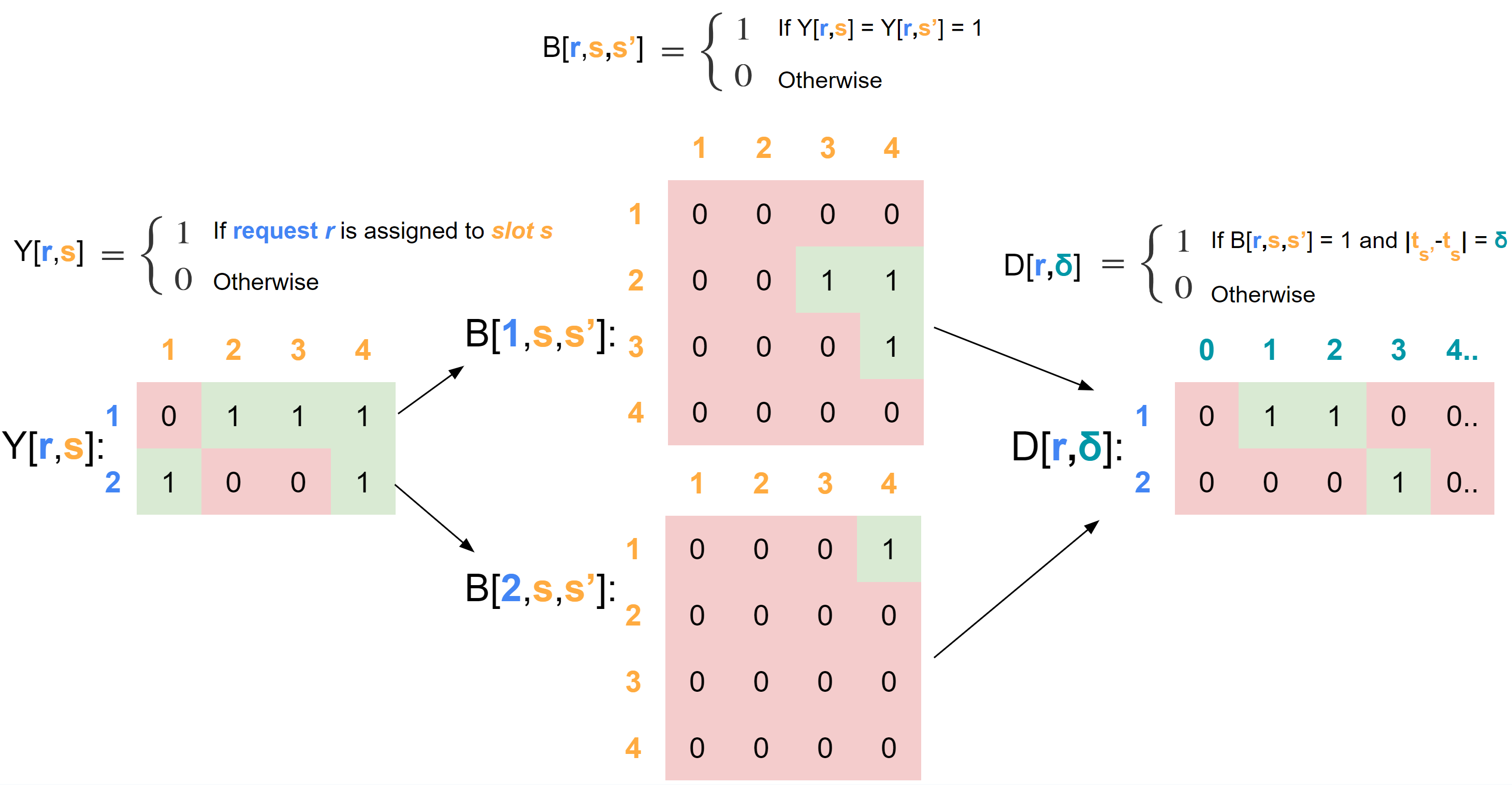}
\caption{{\bf Graphical representation of binary decision variables.} {\bf Left:} $Y_{rs}$ for the first two targets in Figure \ref{fig:yvariable}. To simplify the diagram, we show a scheme with one slot per night where slots occur on sequential nights. Our implementation at Keck uses up to four slots per night and slots are not guaranteed to occur on sequential nights. {\bf Middle:} $B_{rss'}$ specifies any instance when a single request is scheduled into two separate slots, $s$ and $s'$ where $s < s'$. The upper matrix shows how $B_{1,s,s'}$ follows request 1 in $Y_{1,s}$; the lower matrix corresponds to request 2. Entries in $B$ at and below the diagonal are always zero. {\bf Right:} $D_{r,\delta}$ records whether two observations of request $r$ were scheduled $\delta$ days apart.}
\label{fig:variables}
\end{figure*}

Figure~\ref{fig:yvariable} is a schematic of showing how $Y_{rs}$ specifies a  schedule. Figure~\ref{fig:variables} is a schematic showing the relationship between the three binary decision variables listed above.

\subsection{Constraints} \label{sec:semcon}
We achieve the desired observational cadence by enforcing the following constraints in our optimization.

\vspace{0.2cm}

\noindent\textbf{Constraints 1, 2, 3: Establish relationship between $B$ and $Y$}. For each request $r$, the values in $B_{rss'}$ act as \textit{and} statements: values in $B$ are only `turned on' (set to 1) when both corresponding values in $Y$ also contain the value 1. To avoid degenerate constraints, we require $s<s'$. This is accomplished by:

\begin{align}
    B_{rss'}  & \le Y_{rs} \label{eq:bdef1}\\
    B_{rss'}  & \le Y_{rs'} \label{eq:bdef2}\\
    B_{rss' } & \ge Y_{rs} + Y_{rs'} - 1 \label{eq:bdef3}\\
    \forall r & = 1,\ldots,N_r  \nonumber\\    
    \forall s & = 1,\ldots,s'-1  \nonumber\\   
    \forall s'& = 1,\ldots,N_{s} \nonumber
\end{align}
Constraints 1 and 2 ensure that the values in $B_{rss'}$ cannot be 1 when either $Y_{rs}$ or $Y_{rs'}$ is zero. Constraint 3 ensures $B_{rss'} = 1$ when both $Y_{rs}$ and $Y_{rs'}$ are 1.

\noindent\textbf{Constraint 4: Establish relationship between $D$ and $B$.} Values in $D$ are 1 if a pair of scheduled observations exist for a request $r$ at two slots separated by $\delta$. That is, any pair of slot indices in $B$ that satisfy this separation condition will force $D$ to 1:

\begin{equation}
\centering
\begin{aligned}
\label{eq:ddef}
    D_{r\delta} \ge B_{rss'} \quad&\forall r = 1,\ldots,N_r \quad\forall \delta = 0,1,\ldots,\delta_\mathrm{max} \\
    &\forall s = 1,\ldots,s'-1 \,\, \forall s'= 1,\ldots,N_{s} \\
    &\textbf{if} \,\, t_{s'}-t_{s} = \delta
\end{aligned}
\end{equation}

\noindent\textbf{Constraint 5: Targets may only be scheduled when they are visible.} Values in $Y$ must be forced to 0 whenever $V$ holds the value 0.

\begin{equation} 
\begin{aligned}
\label{eq:visible}
    Y_{rs} \le V_{rs} \quad & \forall r = 1,\ldots,N_{r}  \\
    & \forall s = 1,\ldots,N_{s}
\end{aligned}
\end{equation}
If $V_{rs}$ is 1, $Y_{rs}$ may be either 1 or 0. The value is determined as part of the optimization.

\noindent\textbf{Constraint 6: Schedule may not exceed the maximum number of requested observations.} The sum of the values in each row of the $Y_{rs}$ matrix is the scheduled number of observations for request $r$, and should not exceed $n_r^\mathrm{inter}$.

\begin{equation} \label{eq:nobs}
    \sum_{s = 1}^{N_{s}} Y_{rs} \le n_r^\mathrm{inter} \quad \forall r = 1,\ldots,N_r
\end{equation}

\noindent\textbf{Constraint 7: Total exposure times must not exceed length of slot.} We constrain the sum of each column of $Y$ multiplied by the respective exposure time $\tau_{r}^\mathrm{exp}$ and the number of exposures on each night $n_r^\mathrm{intra}$ of each index.

\begin{equation} \label{eq:max}
    \sum_{r = 1}^{N_r} Y_{rs}\tau_{r}^\mathrm{exp}n_{r}^\mathrm{intra} \le I_{s} \quad \forall s = 1,\ldots,N_{s}
\end{equation}
Note that this summation does not give any consideration to slew times or overheads, slightly overfilling some slots. We choose to over-saturate those in the upcoming night, in the next constraint.

\noindent\textbf{Constraint 8: Load the upcoming quarter night(s) with exposures.} Should weather permit ideal observing conditions, slots within the upcoming observing night should be nearly filled with exposures. Let $s^\mathrm{e}$ and $s^\mathrm{\ell}$ be the earliest and latest allocated quarters in the upcoming night, respectively. Let $f$ be the minimum fraction these quarters must be filled, i.e. the lower bound to the column sum.

\begin{equation} \label{eq:min}
    \sum_{r = 1}^{N_r} Y_{rs}\tau_{r}^\mathrm{exp}n_{r}^\mathrm{intra} \ge I_{s}f\quad \forall s = s^\mathrm{e},\ldots,s^\mathrm{\ell}
\end{equation}
We allow the TTP optimization to throw out high slew targets from upcoming nights, if necessary. We note that $s^\mathrm{\ell}$ can arbitrarily be chosen to saturate several nights at once, if one wanted to prepare multiple schedules in advance.

\noindent\textbf{Constraint 9: Ensure minimum specified inter-night spacing of observations}. Finally, we enforce the minimum cadence threshold for each request by restricting the values in $D$, in turn restricting certain pairs of observations.

\begin{equation}
\begin{aligned}\label{eq:mincad}
    D_{r\delta} = 0 \quad&\forall \delta = 0,\ldots,\tau_{r}^\mathrm{inter}-1\\ &\forall r = 1,\ldots,N_r
\end{aligned}
\end{equation}

\subsection{Objective Function} \label{sec:semobj}

The objective of the auto-scheduler is to maximize the number of observations, while penalizing long gaps between observations:

\begin{equation}
    \text{Max}\left( \sum_{r = 1}^{N_r}\sum_{s = 1}^{N_{s}} Y_{rs} - C\sum_{r = 1}^{N_r}\sum_{\delta = 1}^{\delta_\mathrm{max}} \delta D_{r\delta} \right)
\end{equation}
Where $C$ is a small enough constant that the second term is less than unity. We determined $C$ using an upper bound on the value of the second term, assuming that all observations are achieved and any given pair of observations occur a distance $\delta_\mathrm{max}$ apart:

\begin{equation}
    \frac{1}{C} \approx \sum_{1\le r\le N_r,n^\mathrm{inter}_r > 1} {n^\mathrm{inter}_r \choose 2} \delta_\mathrm{max} \approx 10^6
\end{equation}
This ensures that our cadence penalizing term never impedes on the observation of additional targets.

A simple interpretation of the objective is that the optimization algorithm will prioritize a family of solutions to the ILP that have the highest number of completed requests, then search these solutions locally for a schedule that achieves the densest possible sampling for every target. The unconstrained values of $D$ will preferentially be set to 0, since they contribute negatively to the objective function.

We solve this ILP formulation using Gurobi version 10.0.1, a state-of-the-art optimization suite that solves ILP problems using the branch-and-bound algorithm \citep{gurobi}. We process observing requests and allocations using the Python programming language, and generate the ILP model within the Gurobi Python API. In our tests, Gurobi solves the model in ten minutes or less on a modern laptop (see Section \ref{sec:performance} for further details).

\section{Updating the Model Throughout the Semester}
\label{sec:update}

We re-run our algorithm after each observing night by integrating past observing data into our formulation as constraints. The result is a dynamic optimization scheme where a unique semester plan is generated under the new conditions each night.

\subsection{Changes to the Formulation} \label{sec:updcon} 

Updating the environment of the semester problem requires the addition of new constraints and the modifications of others. We first construct a new matrix $P_{rs}$ which contains the ingested observing data from all nights preceding the current night. Recall that $s^\mathrm{e}$ is the first slot in the upcoming night. $P$ will contain entries for all $r$ for times before this value:

\begin{itemize}
    \item $P_{rs}$ = 1 if request $r$ was already observed during slot $s$, and 0 otherwise
\end{itemize}
\noindent\textbf{Constraint 10: At all time values before the present, $\bm{P}$ should dictate the behavior of $\bm{Y}$.}
\begin{equation} \label{eq:past}
\begin{aligned}
    Y_{rs} = P_{rs} \quad & \forall r = 1,\ldots,N_r \\ & \forall s = 1,\ldots,s^\mathrm{e}
\end{aligned}
\end{equation}
In other words, the columns of $Y$ which occurred in the past are included as `variables', but are \textit{not} freely optimized. They are used only to set the initial conditions on cadence for the remainder of the schedule. For this reason, it is important to consider that observers may choose to execute observations that violate constraints imposed on the model. Care must taken to remove the corresponding constraints, or else the values retrieved in $P_{rs}$ may cause the model to become infeasible. 

For example, on a night of notably poor weather conditions, observers might abandon the queue schedule to observe the brightest targets in the request list. Say that one of those chosen targets, $r$, was observed two days ago, but $\tau_{r}^\mathrm{inter}=5$. Observers choose to execute this observation, because it is one of the only targets bright enough for the current conditions. The next day, $P_{rs}$ will contain two observations with $\delta=t_{s'}-t_{s}=2$, which violates Equation \ref{eq:mincad} as a result of Equations \ref{eq:past}, \ref{eq:bdef3}, and \ref{eq:ddef}. This immediately makes the entire optimization infeasible. 

We modify $D$ to not track pairs of slots which have both passed. We redefine the bounds of Equation \ref{eq:ddef} as follows:

\begin{equation}
\centering
\begin{aligned}
\label{eq:ddefnew}
    D_{r\delta} \ge B_{rss'} \quad &\forall r = 1,\ldots,N_r \quad\forall \delta = 0,1,\ldots,\delta_\mathrm{max}\\
    &\forall s = 1,\ldots,s'-1 \quad\forall s' = s^\mathrm{e},\ldots,N_{s} \\  
    & \textbf{if} \,\, t_{s'}-t_{s} = \delta&
\end{aligned}
\end{equation}
For pairs of slots which both occur in the past, $B$ cannot force $D$ to be 1. The model will ignore violations to Equation \ref{eq:mincad} that occurred in the past. The minimum cadence will still be respected for all future values of $Y$.

Next, we restrict the domain of Equations \ref{eq:visible} and \ref{eq:max} to $s = \{s^\mathrm{e},\ldots,N_{s}\}$. The scheduler need only impose these rules on nights which have \textit{not already occurred}, and must accept the values in $P_{rs}$ even if they violate the rules enforced on upcoming nights. The adaptation to \ref{eq:visible} will prevent model infeasibility if a request $r$ was observed during slot $s$, but the value of $V_{rs}$ was computed by the scheduler to be 0. The change to \ref{eq:max} catches the rare case where a past night achieved more exposures than were on the queue schedule, causing the sum of exposures to exceed $I_{s}$ if the constraint is enforced.

\subsection{Re-optimization}

We optimize the semester problem again, with the same objective function as in Section \ref{sec:semobj}. The values in $P_{rs}$  will act as hard constraints, and the optimizer will explore possible plans of action for future slots which maximize the objective. While the semester optimum may be altered after even the first observing night, this dynamic method searches out the best path moving forward. For slots between $s^\mathrm{e}$ and $s^\mathrm{\ell}$, we use the TTP \citep{handleyTTP} to precisely order all the assigned requests and accommodate intra-night cadence constraints, then share the schedule with the observing team.

\section{Implementation at Keck Observatory}
\label{sec:implementation}

This automated queue system is deployed at Keck Observatory by the California Planet Search (CPS) coalition \citep{cps} for the 2023B semester on time allocated for the Keck Planet Finder (KPF). CPS combines a multitude of Doppler search programs, resulting in 1388 observations scattered across 124 different queue quarter nights in 2023B. Within these collective pools is typically a combination of high cadence, low cadence, and single shot observation requests from over a dozen programs.

\subsection{Request Visibility} 
\label{sec:visible}

\begin{figure}
    \centering
    \includegraphics[width=0.4\textwidth]{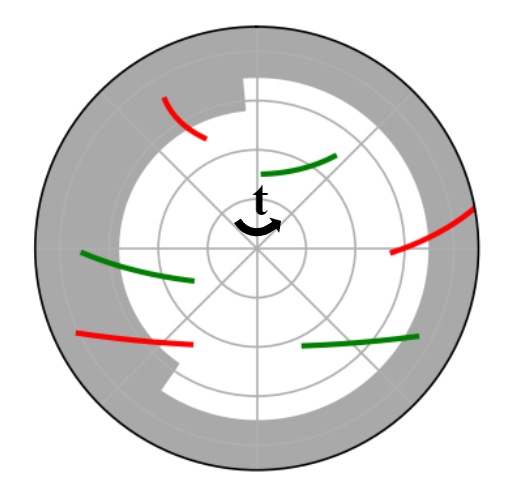}
    \caption{Sample paths in the altitude/azimuth frame at Keck. Targets that would be deemed visible have paths shaded in green as opposed to red. The fraction of time $a$ for which requests must satisfy observing constraints is a free parameter. Here, we show the case where $a=0.5$.}
    \label{fig:visibility}
\end{figure}

Using the start and end Julian Dates of each quarter night, we calculate the duration of each interval $I_{s}$ in minutes, and determine the nominal visibility of each target in the request catalog from the given celestial coordinates. For Keck, we require that the target be above 30 degree elevation, above the Nasymth platform, and more than 30 degrees from the moon. This binary condition is evaluated at every minute within the interval. To reduce the risk of over-constraining the slew optimization described in \cite{handleyTTP} by packing the night too tightly towards the beginning or end of the quarter night, we require that observing conditions be met for some constant $a \times I_{s}$ minutes for $V_{rs}$ to be one (Figure \ref{fig:visibility}), where $a$ can be adjusted as necessary.

It should be noted that it is trivial to make the visibility requirements unique for each request, such as for targets that require lower airmass observations. We compile the subset of nights for which each request is visible at Keck, and construct the matrix $V_{rs}$. 
\subsection{Performance on KPF in 2023B} \label{sec:performance}

We present the first month of progress by our auto-scheduler on the dataset described in Section \ref{sec:inputs}. This semester features fifteen distinct programs with $N_r=204$ unique requests for the pooled queue time of $N_{s}=126$ scattered quarter nights. This totals to 1388 observations to schedule into 317 hours of allocated time. Early in the semester, the problem is largely unconstrained and runs on the order of 10 minutes to obtain a near optimal model. As observing nights pass, the computational load decreases. After one month, it routinely finds an optimal solution in under 3 minutes.

\begin{figure}
    \centering
    \includegraphics[width=\columnwidth]{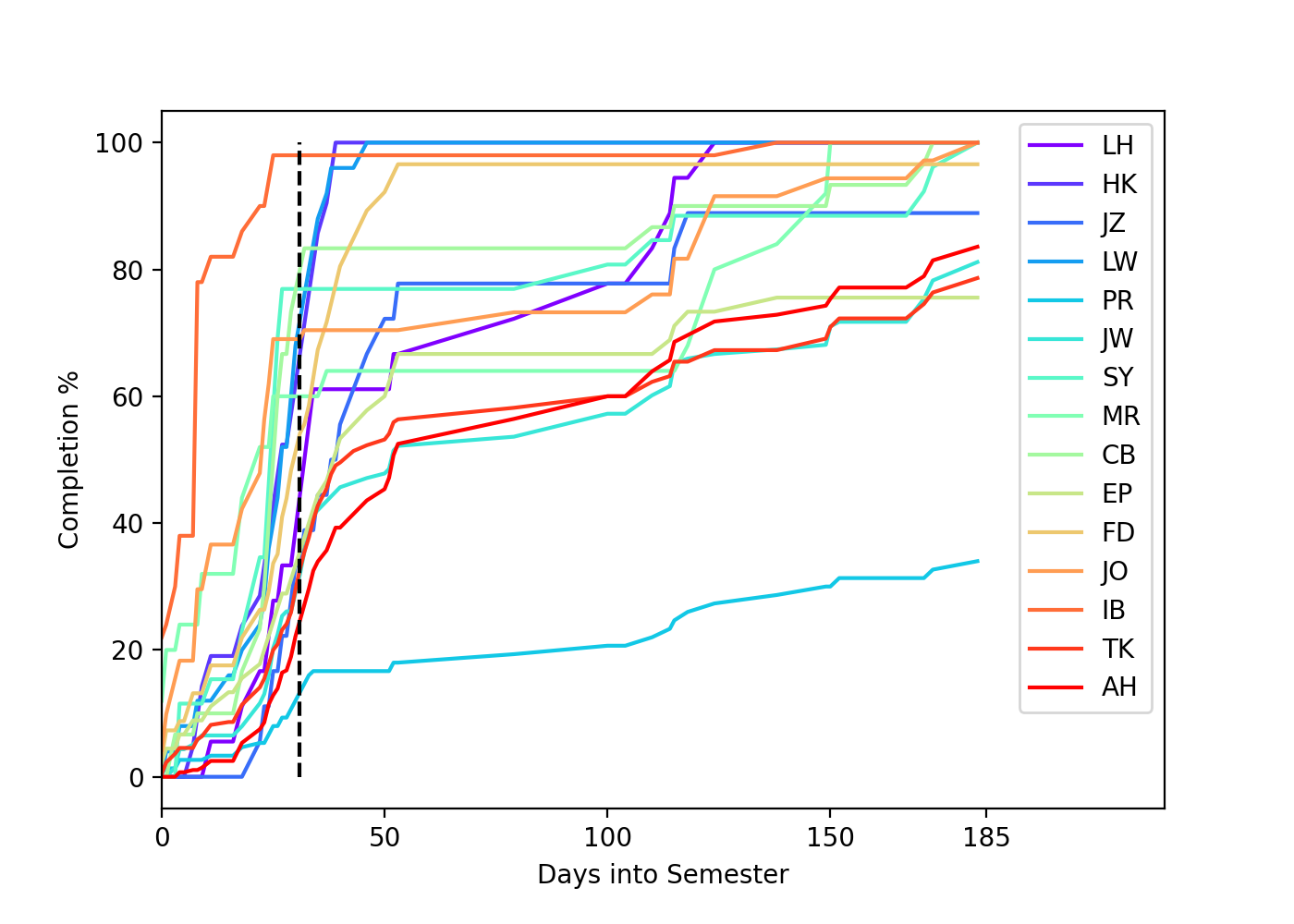}
    \caption{Cumulative progress of each program across the 2023B semester on KPF, as determined by our auto-scheduler. The vertical dashed line indicates the most recent date the scheduler was run (one month of progress).}
    \label{fig:cdfs}
\end{figure}

The scheduler produces various plots after each run. The first shows the completion rate of each program as a function of time. The current and forecasted progress of each program by the scheduler is displayed in Figure \ref{fig:cdfs} subject to all constraints outlined in Section \ref{sec:semcon} and randomized weather sampling. In an ideal world, all programs would reach 100\% at the last day of observations. However, due to mismatches between the distribution of allocated nights determined by the Keck Observatory scheduler (see Section~\ref{sec:background}), and the cadence needs of programs, achieving 100\% completion is not always feasible. 

\begin{figure*}
    \centering
    \includegraphics[width=0.9\textwidth]{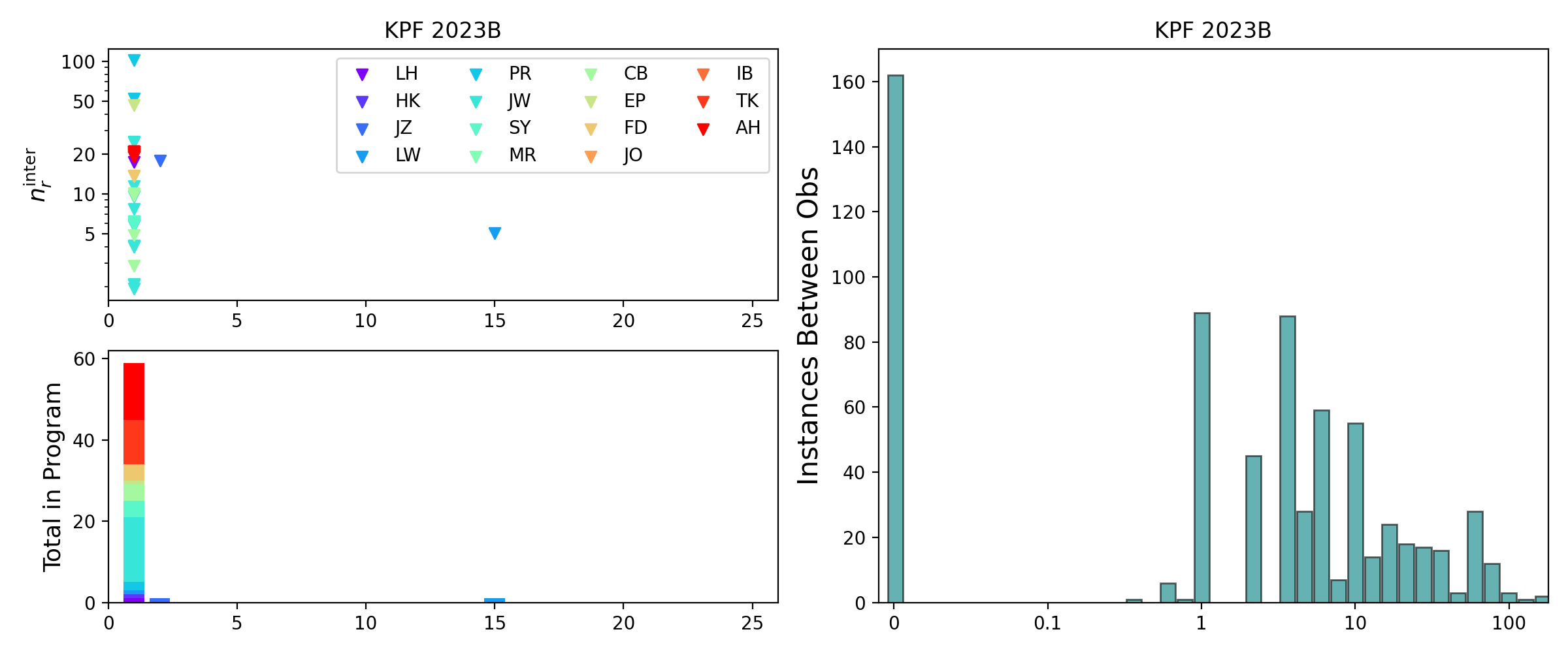}
    \includegraphics[width=0.9\textwidth]{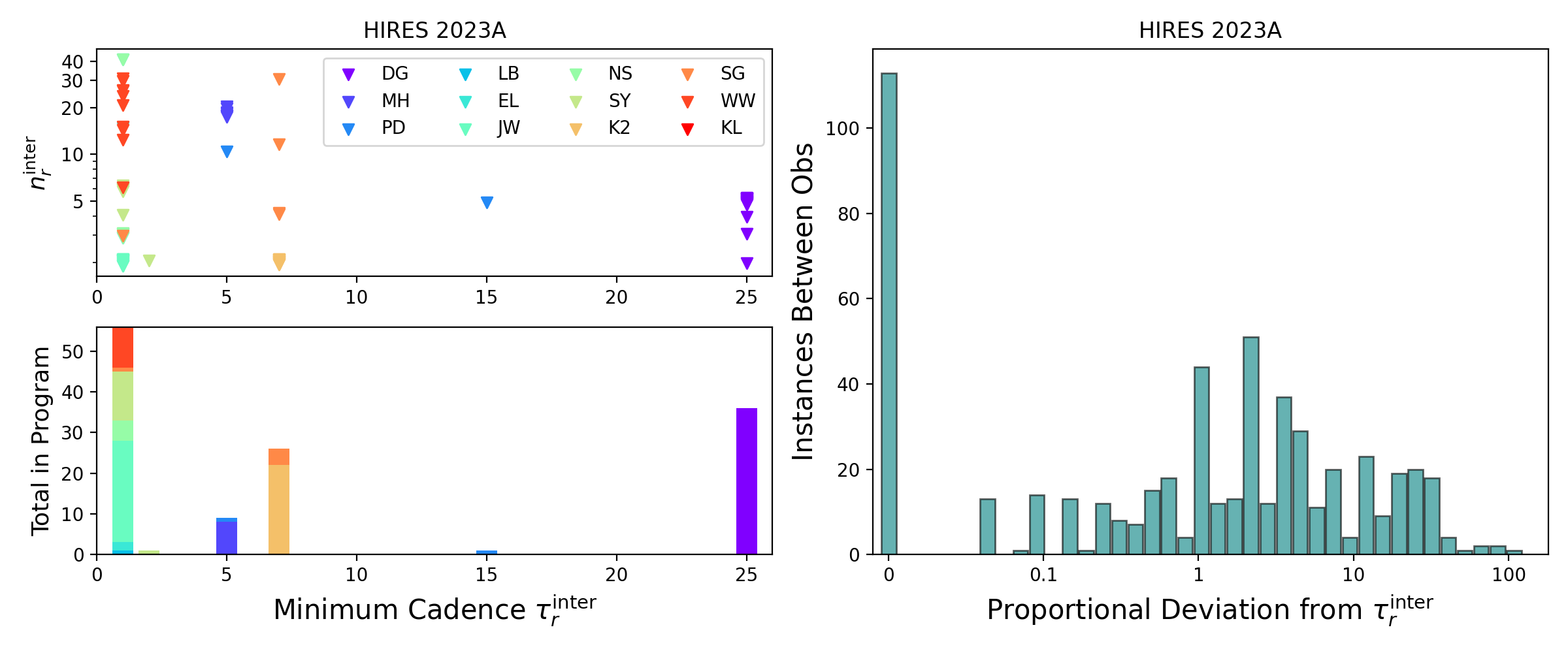}
    \caption{Top left: Distribution of cadenced requests in the $\tau_{r}^\mathrm{inter}$ and $n_r^\mathrm{inter}$ plane for KPF during the 2023B semester, with 5\% induced scatter for visibility. The plot immediately below shows the cumulative number of requests in the respective programs at each cadence. Top right: histogram of how far the simulated schedule deviates proportionally from the ideal cadence for consecutive pairs of observations, across all requests and programs. The bottom three plots are the same as the top, but for HIRES in 2023A.}
    \label{fig:cadence}
\end{figure*}

The Keck Observatory scheduler determines the distribution based on limited information provided in the Keck coversheets, typically consisting of coordinates for the ten highest priority targets. Consequently, the KPF allocations may not align well with the specific cadence requirements of all programs. A summary of the cadence distribution of these requests by program is shown in Figure \ref{fig:cadence}.

\begin{figure*}
    \centering
    \includegraphics[width=\textwidth]{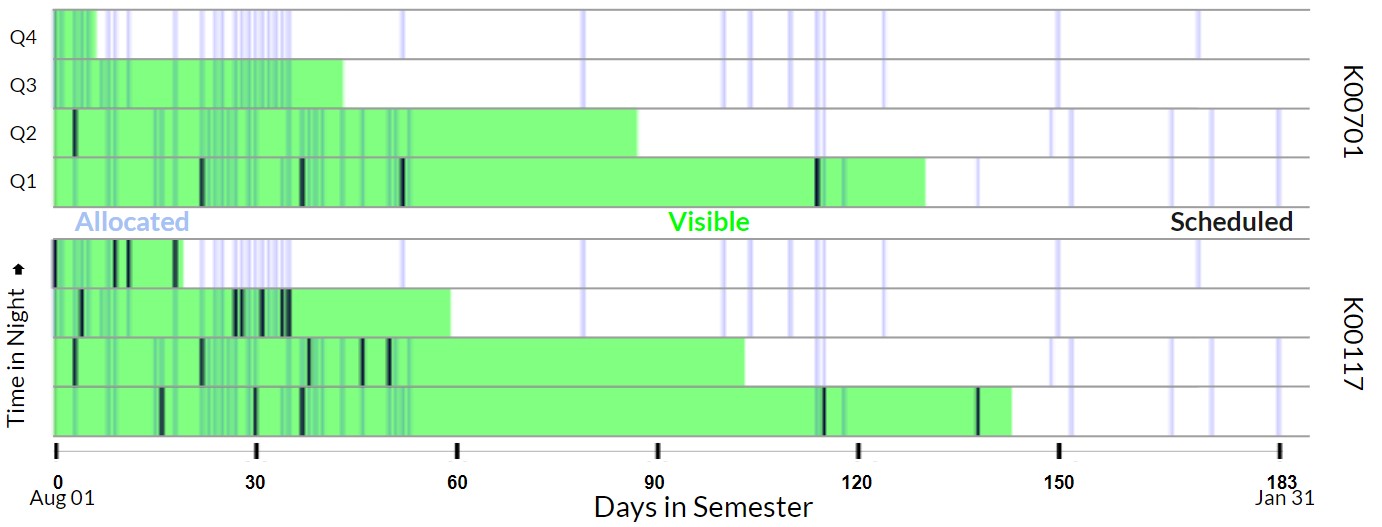}
    \caption{Visualization of the quarter night scheduling problem in 2023B for the two targets K00701 and K00117 (see Table \ref{tab:inputs}) which are low and high cadence requests, respectively. Four rows correspond to each target, with time (quarter) increasing upward until resetting at the next plot. Top plot: visible quarter nights at Keck Observatory (green) throughout each night as a function of the date for K00701. Overplotted are the queue allocated quarter nights (blue) and the scheduled observations (black) for K00701 respecting $n_r^\mathrm{inter} = 5, \tau_{r}^\mathrm{inter} = 15$ as queried from the $Y_{rs}$ variable. Bottom: same as top, but for the target K00117, respecting $n_r^\mathrm{inter} = 20, \tau_{r}^\mathrm{inter} = 1$.}
    \label{fig:programLW}
\end{figure*}

For the entire semester, we also generate target-by-target schedules. Figure \ref{fig:programLW} provides an example for two targets from program LW, illustrating the visibility and scheduling of each target. These plots show when each target is visible, how those periods coincide with allocated time on KPF, and when observations are consequently scheduled. We note that there is a significant difference between the nights available and the nights necessary to observe all targets to completion, which inhibits the progress of some programs. In 2023B, the KPF schedule was heavily front-loaded in the first two months (see Figure \ref{fig:programLW}), which helped some programs achieve high forecasted completions, and hindered others. 

Despite these challenges, eight of the fifteen programs in 2023B are expected to achieve 100\% completion by the end of the semester. Four of the remaining programs are projected to achieve greater than 80\% completion. Ultimately, only one program sits below 75\% completion due to a mismatch in target coordinates and allocated KPF time.

The two targets in the PR program had limited visibility, rising primarily in the second half of 2023B. The target 26965 was requested for $n_r^\mathrm{inter} = 100$ unique nights of observations. The other, 22049, was requested for $n_r^\mathrm{inter} = 50$ unique nights of observations. As computed in $V_{rs}$, both targets were visible for a total of 37 unique nights, bounding the completion to a maximum of less than 50\%.

\begin{comment}
Program completions - - - 
LH 100.0
HK 100.0
JZ 88.9
LW 100.0
PR 34.0
JW 81.2
SY 100.0
MR 100.0
CB 100.0
EP 75.6
FD 96.6
JO 100.0
IB 100.0
TK 78.6
AH 83.6
\end{comment}

The simulated schedule respects all requested minimum cadences, and attempts to schedule each consecutive observation at or near the minimum value. Long gaps between observations are sometimes required due to the distribution of allocated time, but are relatively uncommon. Across all targets, 158 pairs of consecutive observations occurred at exactly the requested cadence, and 57\% of all pairs of observations occur within three days of the minimum cadence. The proportional deviations:

\begin{equation}
    \text{PropDev}(r,s,s') = \frac{t_{s'}-t_{s}-\tau_{r}^\mathrm{inter}}{\tau_{r}^\mathrm{inter}}
\end{equation}
are summarized in the top right of Figure \ref{fig:cadence}.

\subsection{Performance on HIRES in 2023A}

\begin{figure*}
    \centering
    \includegraphics[width=\textwidth]{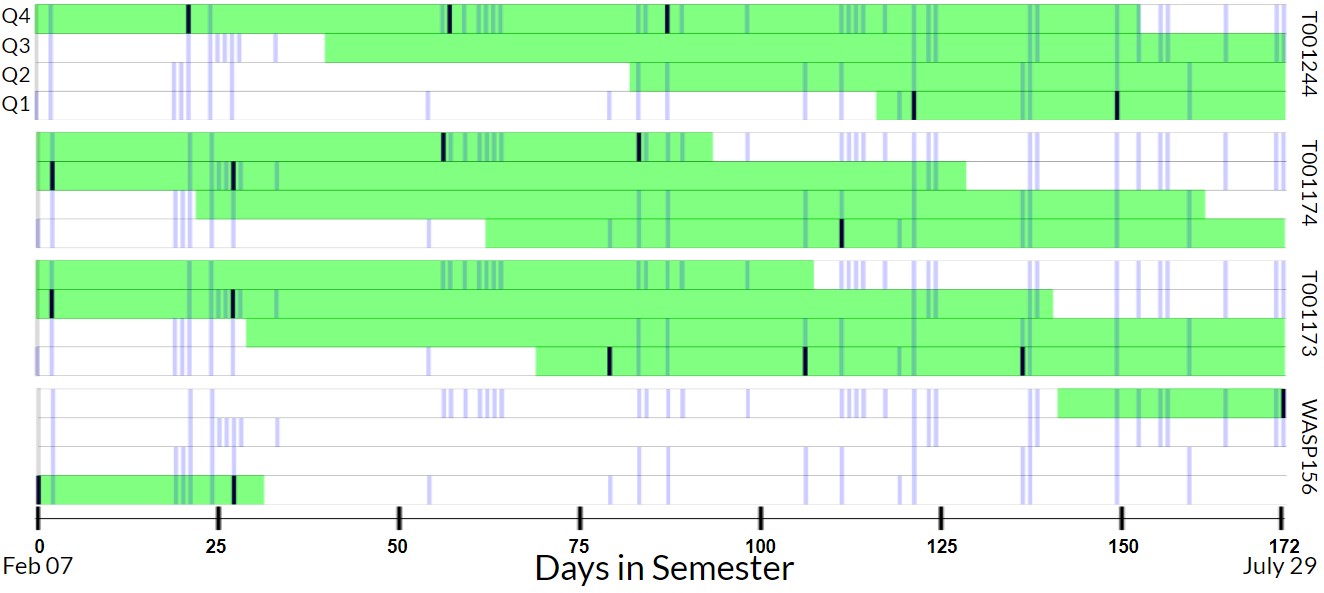}
    \caption{Select targets from our simulation of CPS Doppler observations on the HIRES instrument in 2023A, presented in the same format as Figure \ref{fig:programLW}. Shown are four low cadence requests with $n_r^\mathrm{inter} = 5,\tau_{r}^\mathrm{inter}=25$, which are traditionally difficult to schedule. The last target has severely limited visibility, causing an outlier in the bottom right of Figure \ref{fig:cadence}.}
    \label{fig:programDG}
\end{figure*}

KPF is a newly commissioned instrument and nearly all early science programs in 2023B favor nightly cadence. We wish to demonstrate our algorithm's suitability toward a wider diversity of cadence needs. We further simulate our algorithm's performance on target requests and the allocated time for CPS on Keck/HIRES in 2023A. HIRES in 2023A offers a more mature suite of programs consisting of a mix of high, medium, and low cadence needs, as shown in Figure \ref{fig:cadence}. This data offers a more robust opportunity to test our algorithms performance when cadence is the predominant scheduling constraint. Observations of this type are the most difficult for human schedulers to plan. Of the 543 observational gaps across all requests $r$, 111 occurred at the minimum allowed $\tau_{r}^\mathrm{inter}$ , and 324 occur less than 5 days above $\tau_{r}^\mathrm{inter}$. The median violation of $\tau_{r}^\mathrm{inter}$ is  3 days, with the proportional deviation shown in the bottom right of Figure \ref{fig:cadence}. Additional target-by-target plots of this simulation are shown in Figure \ref{fig:programDG} for a subset of the program DG. Through this simulation, we find that our algorithm is capable of accommodating a mixed portfolio of programs with different cadence needs.

\section{Future Directions} \label{sec:future}
\subsection{Support for Varied Weather Conditions} \label{sec:script}

In principle, the  auto-scheduler operates at sufficient speeds (on the order of 5 minutes on a standard laptop) to support resolving the entire model as weather conditions at Keck evolve. However, these effects can be preemptively mitigated within the scheduling environment by producing multiple observing strategies based upon potential weather conditions for the upcoming night. Furthermore, these environments would benefit substantially from a more sophisticated weather sampling routine than that outlined in Section \ref{sec:weather}. We intend to equip future versions with a Monte Carlo sampler that weathers out nights based on historical WMKO data.

\subsection{Program Equity} \label{sec:equity}

An additional objective of queue systems is to promote equitable instrument access for programs of varying sizes and scientific goals. Some programs may have targets that have limited visibility, causing their scheduling to be disfavored. It may be desirable to equip the framework with variables that track the completion of each program, allowing the objective to homogenize these terms.

The autoscheduler presented here uses a linear objective function. An `equity' term may be added by adding the Average Absolute Deviation (AAD) of program completion may be added. Let $N_{tot}$ be the total number of observations requested across all programs and $N_{p}$ be the number requested by each program $p$, where the list of all programs is $\{1,\ldots,p_{max}\}$. Let $M_{rp}$ be a pre-computed matrix that maps each request $r$ to each program $p$. 

\begin{itemize}
    \item $M_{rp}$ = 1 if the target $r$ is a subset of the program $p$, and 0 otherwise
\end{itemize}
We use an additional variable $A_{p}$ for this formulation, and define this variable with the following constraint:

\begin{itemize}
    \item $A_{p}$ is the normalized (percentage) absolute deviation of each program from completion
\end{itemize}

\begin{align}
    A_{p} = \left | \frac{\sum_{r = 1}^{N_r}\sum_{s = 1}^{N_{s}} M_{rp}Y_{rs}}{N_{p}} - \frac{\sum_{r=1}^{N_r}\sum_{s=1}^{N_{s}} Y_{rs}}{N_{tot}} \right | \\  \forall p =1,\ldots,p_{max} \nonumber
\end{align}
The objective in section \ref{sec:semobj} can be modified to include the sum of these deviations for all programs times a scalar. However, solving with 3 objective terms reduces the interpretability of the ILP framework, since the scalar terms in the objective function need to be fine tuned by hand to find a balance of program equity and observational cadence. Setting a maximum acceptable deviation using the terms in $A_{p}$ as a \textit{constraint} on the model avoids this ambiguity.

\section{Conclusion}

In summary, we have developed an automated scheduler to manage the execution of approximately $\sim$1000 Doppler observations during a single semester at the Keck Observatory. Our on-sky progress with KPF in 2023B demonstrates that the system consistently achieves high completion rates across various programs while adhering to inter-night and intra-night cadence specifications. Additionally, our simulations on HIRES in 2023A showcase the system's proficiency in handling diverse programs with varying cadence requirements, ensuring a significant portion of observations occur at the requested frequency.

Our scheduling system operates efficiently, capable of scheduling an entire semester within a few minutes. This not only saves considerable time for human schedulers but also provides a long-range outlook, alleviating CPS participating scientists from having to manually request targets for specific nights. The subsequent Traveling Telescope Problem optimization minimizes instrument downtime during observing runs, offering precise execution timing for each observation.

This algorithmic framework can be extended for use in any survey that involves collecting a large number of observations over numerous distinct observing intervals, particularly when the cadence of observations is a critical factor. As the coming decades witness an increase in such endeavors, scheduling algorithms equipped to manage these challenges will become indispensable in achieving scientific objectives. Simulations employing this algorithm or variants can enhance our understanding of how allocated time distribution impacts different programs or surveys and how target selection can be optimized at ground-based observatories. 

\begin{acknowledgments}
L.H., E.A.P., and J.L. acknowledge support from the Heising-Simons Foundation Grant \#2022-3832. V.V.M. acknowledges support from the UCLA Anderson School of Management. Additionally, the authors wish to recognize and acknowledge the very significant cultural role and reverence that the summit of Maunakea has within the indigenous Hawaiian community. We are most fortunate to have the opportunity to conduct observations from this mountain.

\end{acknowledgments}



\software{astropy \citep{2013A&A...558A..33A,2018AJ....156..123A}, numpy \citep{numpy}, pandas \citep{pandas}, gurobi \citep{gurobi}, matplotlib \citep{matplotlib}}


\appendix
\restartappendixnumbering
\section{Variables} \label{sec:variables}

Table~\ref{table:variables} lists the variables used in all preceding sections of this paper, along with their first usage:
\begin{table*}
    \centering
    \caption{Symbols Used}
     \label{table:variables}
    \begin{tabular}{|c|l|c|}
     \hline
     \multicolumn{1}{|c|}{\textbf{Symbol}}
      & \multicolumn{1}{c|}{\textbf{Definition}} & \multicolumn{1}{c|}{\textbf{Section}}\\
     \hline
     $a$ & Minimum portion of an interval for which a request must satisfy visibility constraints to be scheduled & \ref{sec:visible}\\
     $A_{p}$ & Continuous variable indicating the absolute deviation of completion percentage of program $p$ from & \ref{sec:equity}\\
     & the total semester completion percentage & \\
     $B_{rss'}$ & Binary variable that indicates whether request $r$ is assigned to both slots $s,s'$ concurrently & \ref{sec:semvar}\\
     $C$ & Small constant used in the objective function to ensure the scheduling of additional observations is  & \ref{sec:semobj} \\
     & always prioritized & \\
     $D_{r\delta}$ & Binary variable that indicates whether request $r$ has multiple observations separated by an interval & \ref{sec:semvar}\\
     & of $\delta$ days & \\
     $f$ & Minimum fraction of time for which the upcoming nights should be filled with exposures & \ref{sec:semcon}\\
     $I_{s}$ & Duration of the slot $s$ & \ref{sec:visible} \\
     $M_{rp}$ & Pre-computed program map. Holds 1 if request $r$ is a subset of program $p$, 0 otherwise & \ref{sec:future} \\
     $n_r^\mathrm{inter}$ & Maximum number of unique nights of observation over the semester for request $r$ & \ref{sec:inputs} \\
     $n_r^\mathrm{intra}$ & Maximum number of exposures per scheduled night for request $r$ & \ref{sec:inputs} \\
     $N_{p}$ & Total observations requested by the program $p$ & \ref{sec:equity}\\
     $N_r$ & Total number of unique requests & \ref{sec:inputs} \\
     $N_s$ & Total number of allocated slots & \ref{sec:inputs} \\
     $N_{tot}$ & Total observations requested in the semester problem across all programs & \ref{sec:equity} \\
     $P_{rs}$ & Past observation matrix. Holds 1 if request $r$ has already been observed during the & \ref{sec:updcon} \\
     & slot $s$, 0 otherwise & \\
     $p$ & Program index & \ref{sec:equity}\\
     $p_{max}$ & Number of unique programs & \ref{sec:equity}\\
     $r$ & Request index & \ref{sec:inputs} \\
     $s,s'$ & Slot indices & \ref{sec:inputs} \\
     $s^\mathrm{e},s^\mathrm{\ell}$ & Earliest and latest slots indices which exist in the immediately upcoming night, respectively & \ref{sec:semcon} \\
     $t_s$ & Calendar date of the slot $s$ relative to the first slot allocated (days) & \ref{sec:inputs} \\
     $V_{rs}$ & Visibility matrix. Holds 1 if request $r$ is visible during the slot $s$, 0 otherwise & \ref{sec:visible} \\
     $Y_{rs}$ & Binary variable that holds the scheduling state of request $r$ to slot $s$. Holds 1 if request $r$ is & \ref{sec:semvar}\\
     & assigned to quarter night $s$, 0 otherwise &\\
     $\delta$ & Time difference (days) & \ref{sec:inputs}\\
     $\delta_\mathrm{max}$ & Calendar difference between the final and first slot in the semester (days) & \ref{sec:inputs}\\
      $\tau_{r}^\mathrm{exp}$ & Nominal exposure time of request $r$ (seconds) & \ref{sec:inputs} \\
     $\tau_{r}^\mathrm{inter}$  & Minimum inter-night cadence between observations of the same request $r$ (days) & \ref{sec:inputs}\\
     $\tau_{r}^\mathrm{intra}$ & Minimum intra-night cadence between observations of the same request $r$ (hours) & \ref{sec:inputs}\\
     \hline
    \end{tabular}
\end{table*}


\bibliography{bibliography}{}
\bibliographystyle{aasjournal}



\end{document}